\documentclass[a4paper, 10 pt, conference]{ieeeconf}

\IEEEoverridecommandlockouts 
\usepackage[USenglish]{babel} 
\usepackage[T1]{fontenc}
\usepackage[ansinew]{inputenc}
\usepackage[noadjust]{cite}
\usepackage{lmodern} 
\usepackage{graphicx} 
\usepackage{float}

\usepackage{amsmath}
\usepackage{amsfonts}
\usepackage{amssymb}
\usepackage{bm}
\usepackage{url}
\usepackage{hyperref}
\usepackage{booktabs}
\usepackage[font=small,labelfont=bf]{caption}

\newcommand{\Ncov}{N_{\text{cov}}}

\newcommand{\Ns}{N_{\text{sens}}}
\newcommand{\Nprio}{N_{\text{prio}}}
\newcommand{\Nroad}{N_{\text{road}}}
\newcommand{\pdiv}{p_{\text{div}}}
\newcommand{\pmut}{p_{\text{mut}}}
\newcommand{\pcross}{p_{\text{cross}}}
\newcommand{\cov}{c_{\text{}}}
\newcommand{\ceff}{c_{\text{eff}}}
\newcommand{\gs}{l_{\text{grid}}}
\newcommand{\m}{\text{m}}

\renewcommand{\baselinestretch}{0.97} 

\begin{document}

\title{\bf Optimized sensor placement for dependable roadside infrastructures}

\author{
\authorblockN{Florian Geissler{$^*$}\thanks{{$^*$} Email: florian.geissler@intel.com} and Ralf Gr\"afe}\\
\authorblockA{Dependability Research Lab, 
Intel Labs Europe\\
85579 Neubiberg, Germany
} }
\date{\today}
\maketitle

\begin{abstract}
We present a multi-stage optimization method for efficient sensor deployment in traffic surveillance scenarios.
Based on a genetic optimization scheme, our algorithm places an optimal number of roadside sensors to obtain full road coverage in the presence of obstacles and dynamic occlusions. 
The efficiency of the procedure is demonstrated for selected, realistic road sections.
Our analysis helps to leverage the economic feasibility of distributed infrastructure sensor networks with high perception quality. 
\end{abstract}

\section{Introduction}

Intelligent transportation systems pose one of the key challenges for our modern society. 
To keep pace with the ever increasing demand for safe and fast mobility, while at the same time dealing with the limitations of available space and traffic volume, future transportation requires a higher level of i) dependability (including safety), ii) efficiency and iii) affordability.
This is first and foremost relevant for road traffic in densely populated areas.
Automated vehicles (AV) are expected to facilitate all of these aspects, and the effort towards a large-scale implementation is ubiquitous.

One approach to automation is an empowerment of the vehicle's individual capabilities, i.e. an increasing level of autonomy. On the other hand,
traffic agents benefit from the collaboration with other agents (vehicle-to-vehicle, V2V) or a roadside infrastructure (vehicle-to-infrastructure, V2I). The latter typically consists of a network of sensors for object detection, and a wireless communication path to the vehicles. 
Complementing the limitations of a vehicle-centric approach, a roadside infrastructure contributes to the above challenges in the following ways: i) It provides additional, safety-relevant information about the environment, that vehicles may not be able to perceive by means of on-board sensing only, e.g. due to occlusions or limited range. ii) It acts as a well-informed authority for traffic coordination, and therefore provides a handle to enhance traffic flow. iii) While the costs of setting up an infrastructure can represent an initial barrier, the long-term saving effects seem very promising. First of all, centralized computational resources in the infrastructure can supersede in-vehicle devices, which are less powerful and, in relative terms, more costly.
The potential of infrastructure support to AVs has long been recognized. For example, the supervision of critical hotspots such as intersections has been a main focus. Cameras and communicating traffic lights improve the road safety and traffic flow at smart intersections \cite{Fuerstenberg2005, Leader2004}.
A variety of V2I applications -- including blind spot detection, signal phase control, or road condition monitoring -- are investigated in complex test environments such as Mcity \cite{Fayta2015} or CETRAN \cite{Quek2016}. Car manufacturers are ready to keep up with this technology \cite{Audi2016}.

A natural extension to such an infrastructure concept is the deployment of sensor networks not only to monitor selected points of high risk, but for a comprehensive and seamless coverage of the road. Pushing the trend towards reduced vehicle autonomy, the infrastructure can be enabled to take over more complex tasks such as cooperative trajectory planning, or even remote control. 
This requires dependable perception qualities of the sensor network, first of all complete sensing information, which makes optimal sensor placement a key prerequisite.
A roadside infrastructure for the seamless surveillance of individual roads was for instance studied as part of the European SAFESPOT project \cite{Safespot2006}, using camera and laser sensors, whereas the German KoRA9 project \cite{BMVI2019} targets highway monitoring based on radar sensors. In both cases, a simplistic linear network topology is used.
With a view to more complex, dynamic environments -- such as smart cities, campuses, or automated parking systems -- we see the need for improved sensor deployment strategies. 

The article is structured as follows: Sec.~\ref{sec:relatedwork} discusses related literature before our improved sensor deployment method is presented in the Sec.~\ref{sec:model}. We demonstrate the efficiency of the algorithm for selected realistic scenarios in Sec.~\ref{sec:results}. Finally, we conclude in Sec.~\ref{sec:summary}.

\begin{figure*}[htb]
\centering
\begin{minipage}{1\textwidth}
  \centering
  \raisebox{-0.5\height}{\includegraphics[height=5.3cm]{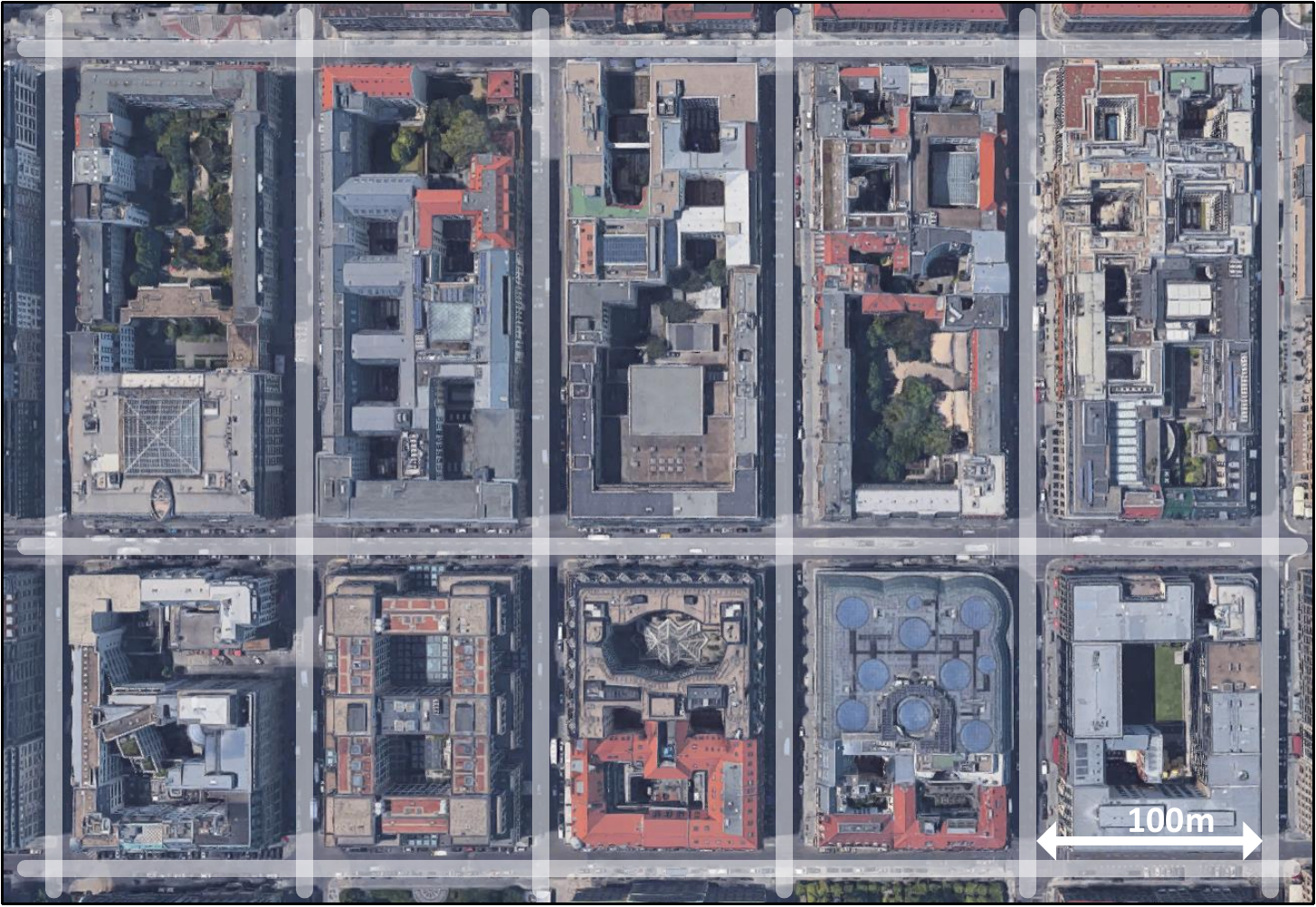}}
  \raisebox{-0.55\height}{\includegraphics[height=6.8cm]{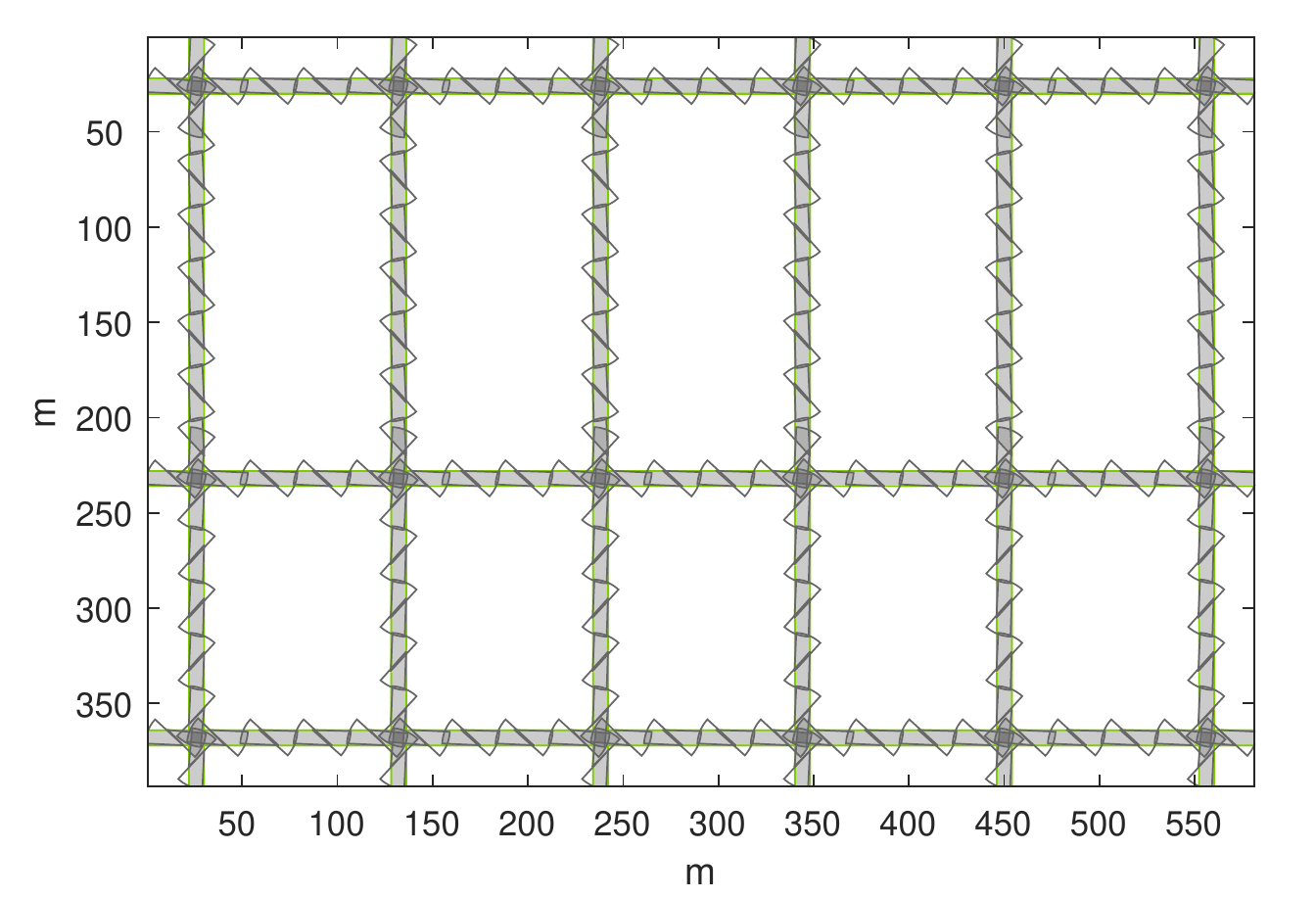}}
\end{minipage}
\caption{Optimized placement of camera sensors in a section of Berlin city. (Left hand side) Satellite view of the scene, as taken from Google$^{\circledR}$ maps \cite{GM2019}. (Right hand side) Infrastructure deployment of sensors with a sensing field of $r=20\m$, $\omega=40^{\circ}$. Our algorithm achieves full coverage ($\cov=1$) at high efficiency ($\ceff=0.64$), spending $312$ sensors. For more details see Sec.~\ref{sec:scaling}.}
\label{fig:city}
\end{figure*}

\section{Related work}
\label{sec:relatedwork}

In the last decades, area coverage problems have found a natural interpretation and continuation in terms of the placement of nodes in spatially distributed (wireless) sensor networks (WSN). Depending on the sensor model, a perfect tessellation is in general not possible, and a trade-off between coverage level and sensor costs has to be determined. Whereas omni-directional models characterize sensor nodes only by their range \cite{Zhu2012}, the degrees of freedom of the field of view (FoV) and the working direction angle add to the complexity of the coverage problem in directional sensor networks (DSN) \cite{AmacGuvensan2011}. 
%
A well-studied scenario is the placement of camera sensors to cover a confined three-dimensional environment such as the interior of a building \cite{VanDenHengel2009, Indu2009}. 
Other work focuses on the seamless coverage of an extended open terrain (for example a battlefield) \cite{Yoon2013, Zhang2016}.
While in these scenarios the regions of sensor deployment and surveillance coincide, less attention has been dedicated to setups with reduced or even zero overlap. Importantly, different boundary conditions for the two respective cases impact the choice of applicable optimization methods. 
The requirements for complete coverage of a straight road segment with omni-directional sensors along the roadside is analyzed in \cite{Cheng2010}. 
Another line of research is concerned with the placement of sensors at strategic locations in large-scale road networks, see for instance in \cite{Field2006}. Here, the objective is the surveillance and control of the statistic traffic flow, not seamless road coverage.

Due to the typically large search space and the density of local extrema of the cost function, area coverage problems challenge traditional optimization schemes. 
Methods found in the literature range for instance from connected subgraphs \cite{Tao2006}, a generalized virtual field approach \cite{Cheng2007}, and repulsive potential fields \cite{Zhao2009} to Voronoi diagrams \cite{Zhang2016}. 
To address the DSN placement problem in a numerical way, genetic algorithms (GA) \cite{Holland75} have emerged as a particularly useful tool \cite{Indu2009, Yoon2013}. 
%
Furthermore, meaningful design guidelines for a roadside infrastructure have to include objects dynamics, 
since vehicle traffic on a specific street segment can constitute a significant sensing barrier for behind street segments. 
The probability of occlusions can be reduced for instance by enforcing multiple coverage in specified priority regions \cite{Indu2009}. Alternatively, the interception of likely vehicle paths can act as an additional objective for sensor placement (similar to e.g. \cite{Ferrari2006, Field2006}).

The present article goes beyond current art in the following ways (see Sec.~\ref{sec:model}):
\begin{itemize}
\item We study the coverage of arbitrary road geometries by a DSN of variable sensor type. The region for non-intrusive sensor placement (the roadside) has no overlap with the area to be covered (the road). 

\item The harnessed genetic algorithm implements a customized crossover operation to breed offspring in a guided way.

\item We not only include the possibility of line-of-sight (LoS) obstacles and priority areas, but further introduce semi-transparent obstacles as a new feature to account for object dynamics. 

\item We make use of symmetries of the underlying environment to further improve automated sensor placement. 

\item We demonstrate a bottom-up approach to achieve near-optimal sensor placement on largely extended maps, by efficiently interfacing small-scale solutions.

\end{itemize}


\section{Model}
\label{sec:model}

\begin{figure*}[!tb] 
\includegraphics[width=1\textwidth]{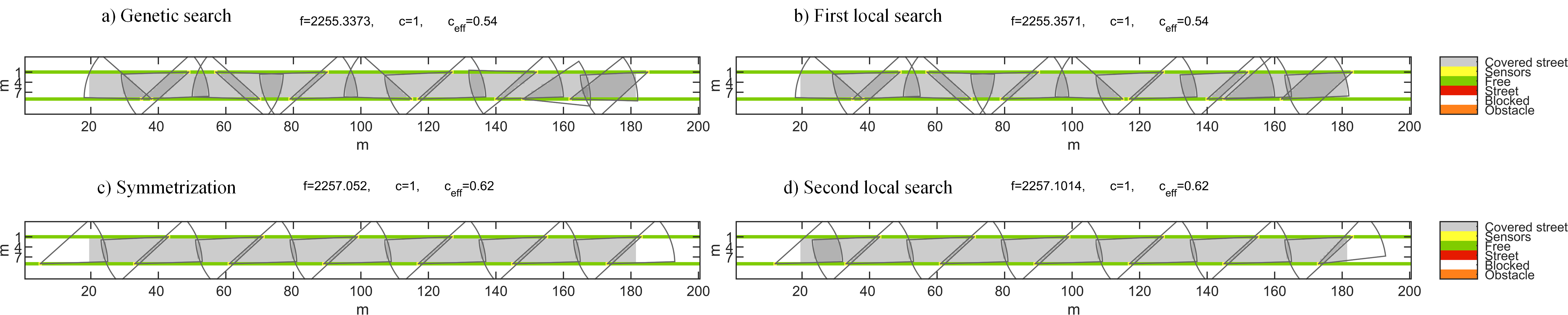}
\caption{Illustration of the optimization procedure for the example of a straight road segment with two lanes, covered by camera sensors ($r=20\m$, $\omega=40^{\circ}$). The outcome of the genetic evolution (a) is refined by a greedy local search (b). Subsequently, symmetrization (c) identifies a favorable translational symmetry of $28\m$. A final local search only adjusts the two road ends in this case (d).
Grid resolution is $\gs=1\m$, and we assume a lane width of $3.5 \text{m}$ throughout the article.}
\label{fig:normalroad}
\end{figure*}

The elements and routines of the model are discussed individually below.
Our optimization procedure follows the general structure of a genetic algorithm, as e.g. outlined in \cite{Srinivas94}.
An illustration is given in Fig.~\ref{fig:normalroad}. 

\subsection{Environment model and sensor encoding}

The environmental scene is defined as a two-dimensional grid with variable cell size. Each square is assigned one of the five tags \textit{obstacle}, \textit{blocked}, \textit{street}, \textit{free}, \textit{sensor}. 
Here, an {\it obstacle} blocks the line of sight of other sensors in range (e.g. building walls, vegetation etc.), while a {\it blocked} square does not interfere with sensors (e.g. sidewalks). The set of all {\it street} cells form the area to be covered by sensors, and the {\it free} squares define available positions for sensor placement. Once a sensor was placed, the cell obtains the tag {\it sensor}, which precludes the further positioning of sensors at the same position.
Two additional features address the issue of dynamic occlusions in the sensor placement process. First, a subset of the {\it street} squares can be assigned the {\it priority} objective of being covered at least twice. 
On the other hand, we allow for a hybrid cell type of {\it street} and {\it obstacle} with a variable degree of transparency, meaning that a corresponding number of randomly selected grid cells in the shadow of the obstacle is occluded. The transparency value can be set to reflect the expected traffic density (see Sec.~\ref{sec:scenarios}). 

A sensor object $s$ is defined by the tuple $s=\{r, \omega; x,y, \phi\}$,
where $r$ is the maximum sensing range, $\omega$ the horizontal field of view (FoV), $x$ and $y$ the integer grid coordinates of the sensor location, and $\phi$ the sensor orientation angle relative to the $x$-axis. The parameters $r$ and $\omega$ can be adapted to model different sensor types such as camera, radar or lidar sensors. In this article, we assume that all sensors deployed in a given scenario are homogeneous, and $r$ and $\omega$ are thus not subject to the optimization procedure. The continuum of orientation angles is reduced to the finite number of viewing angles from a selected location to all existing street cells of the scene. A sensor is assumed to exhibit uniform detection capabilities across its FoV.
The variable sensor parameters represent the genes of the genetic algorithm, while a given set of sensors forms a chromosome, or solution. 

\subsection{Fitness function}

The fitness function $f$ evaluates the quality of a solution. Let us by $\Ncov(n)$ define the number of street cells that are covered {\it at least} $n$ times by a respective sensor network configuration (where we consider a street cell as \textit{covered} by a sensor if the grid center is in the respective sensor's FoV).
With that, $f$ is chosen to be 
\begin{align}
\label{eq:fitness}
f=\alpha \Ncov(1) + \beta \Nprio - \gamma \Ns + \delta \sum_{n=2}^{\Ns} \frac{\Ncov(n)}{n-1}.
\end{align}
Here, $\Nprio$ represents the number of street cells whose additional priority constraints were satisfied, and $\Ns$ is the total number of deployed sensors. The first and second term in (\ref{eq:fitness}) attribute a reward in case a street cell is covered at all, or satisfies its predetermined priority, respectively. The third term penalizes the use of additional sensors, while the last term rewards the overlap of the FoVs of multiple sensors. Note that the latter optimizes the efficiency of the solution, as it tries to avoid a waste of sensing space, if full coverage of the scene is already achieved. 
To promote a more homogeneous coverage, overlap of a higher degree is assigned a slightly reduced reward.
The weighting factors $\alpha, \beta, \gamma$, and $\delta$ determine the hierarchy of the various objectives of the optimization procedure.
For dependable surveillance, we stipulate the boundary conditions $\beta= \alpha-\delta$ (to avoid double counting), $\alpha > \gamma$, and $\gamma > \delta$. In particular, to avoid extra overlap at the cost of additional sensors, we here use $\{\alpha, \beta, \gamma, \delta\}=\{2\Nroad, 2\Nroad-1, \Nroad, 1\}$, where $\Nroad$ is the number of street cells in the scenario.

\subsection{Selection, crossover and mutation} 
\label{sec:crossover}

Starting from an initial population size of $N$, parents are randomly paired in each generation to breed one child chromosome, with a crossover probability $\pcross$. Subsequently, the population is resized by selecting $N$ solutions from the total pool of $N(1+\pcross/2)$ chromosomes in the following way: The fittest ten percent of the population are directly transferred to the next generation, while diversity is maintained by an injecting rate of $\pdiv N$ new chromosomes to the mating pool. The remaining slots are filled by a roulette wheel selection scheme \cite{Holland75}. 
We further include elitism of the best chromosome of a generation. 

An intuitive crossover approach was to swap a certain number of sensors from two parent solutions \cite{VanDenHengel2009}. We find, however, that this does not provide a very efficient optimization path for our variation of the coverage problem. Instead, we use a more guided crossover function, in the form of sequential gene ranking. 
To crossbreed a new offspring chromosome, the individual genes of the two parent solutions are ranked in terms of i) the number of street squares they cover, given the current orientation and position, and ii) the number of street squares that are in range, given the current position. The latter is of interest because sensors with more street cells in their vicinity have a higher chance to end up in a valuable configuration, after a subsequent rotation. This parameter acts as a secondary decision criterion to break ties with respect to the primary criterion. 
The gene with the best ranking is carried over to the offspring chromosome, and the respective portion of covered street cells is removed from the scene. The crossover operation terminates if no parent genes with non-zero coverage contribution remain.

We apply a Gaussian mutation scheme \cite{Deb2001} to the pool of all solutions, which disturbs an average portion of $\pmut$ percent of the population. For a given gene, the equally likely options of mutation are a modification of the position, the orientation angle, or the deletion of the respective sensor. Furthermore, we allow for a finite chance that mutation adds a random gene to the chromosome.

\subsection{Termination threshold and success metric}

The optimization procedure is stopped if the fitness value does not change anymore over a period of five consecutive generations. 
To give a more intuitive understanding of the quality of a configuration, we define in addition to (\ref{eq:fitness}) the success metric, 
\begin{align}
\label{eq:metricSimple1}
\cov &= \frac{\Ncov(1)}{\Ncov(0)} \propto \Ncov(1), \\
\label{eq:metricSimple2}
\ceff &= \frac{\Ncov(0) \gs^2}{\Ns r^2 \omega/2} \propto \Ns^{-1}.
\end{align}
Here, $\gs$ is the grid cell length. While $0 \leq \cov \leq 1$ quantifies the covered portion of the area of interest, $\ceff$ relates to the average density of the sensor configuration with respect to the street space. The closer the solution gets to $\ceff=1$, the more efficient is the solution.

\subsection{Greedy local search}

GAs are well-suited to find solutions of high fitness within a large search space, however, their nature makes the identification of a global optimum in general improbable. We therefore apply a subsequent greedy search optimization, to further refine the optimal sensor configuration. The search follows the steepest ascent of the global fitness function, as determined by local variations of each individual sensor of the ensemble. Explicitly, for each sensor we vary i) the location, up to any of the twelve next-nearest neighbors of the grid, if available for sensor positioning, ii) the working direction, checking the ten next best discrete angles for nearest neighbors. The local search tests as well if it is favorable to iii) eliminate the respective sensor.

\subsection{Symmetrization}

Depending on the form of the sensing field, the topography of the sensor placement region etc., the placement procedure may naturally reproduce characteristic symmetry patterns of the underlying map. In anticipation of this effect, the appropriate symmetrization of candidate solutions can
help the success of the method. However, while symmetry patterns are rather intuitive to the human eye, typical optimization algorithms are agnostic of this 
feature. We here choose the following approach: A given chromosome is first augmented with symmetry seeds, meaning that all operations of a symmetry group are subsequently applied, and the respective sensors are added to the ensemble.
Next, an elimination operator seeks to pick the best, and at the same time most symmetry-compliant sensors from the augmented solution. This operator is very similar to a self-crossover operation discussed in \ref{sec:crossover}, just that the ranking method is modified. We now rank genes by i) the number of times a sensor appears in the augmented solution, and ii) the number of street cells covered. Hereby, the operator picks the most symmetry-compliant seed from the full sensor ensemble in the first iteration, and from thereon refers to the offspring solution itself to pursue a started symmetry pattern.
We find that the optimization procedure is highly impacted by the number of pattern breaks. A pattern break occurs if no available sensor complies with the existing symmetry, but there are still street cells to be covered. By restricting the allowed number of pattern breaks, as estimated from the scene, the optimization procedure can be facilitated. 
After the symmetrization step, another local search helps to remove redundant sensors.
For translation-symmetric maps, a subroutine examines the optimal translation vector.


\begin{figure*}[htb] 
\includegraphics[width=1\textwidth]{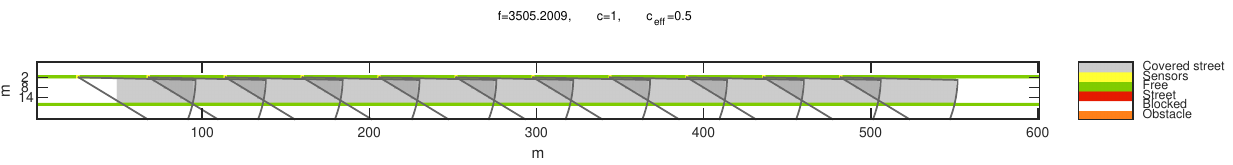}
\caption{Four-lane motorway covered by mid-range radar sensors ($r=70\m$, $\omega=30^{\circ}$). The rightmost (lowest) lane is modeled here as an $80\%$-opaque obstacle to reflect the higher traffic density compared with the left (higher) lanes. The grid resolution is $\gs=2\m$. Finite size effects are reduced by cutting off both ends of the highway, compared to the placement region.}
\label{fig:highway}
\end{figure*}

\begin{figure}[htb]
\centering
\includegraphics[width=.5\textwidth]{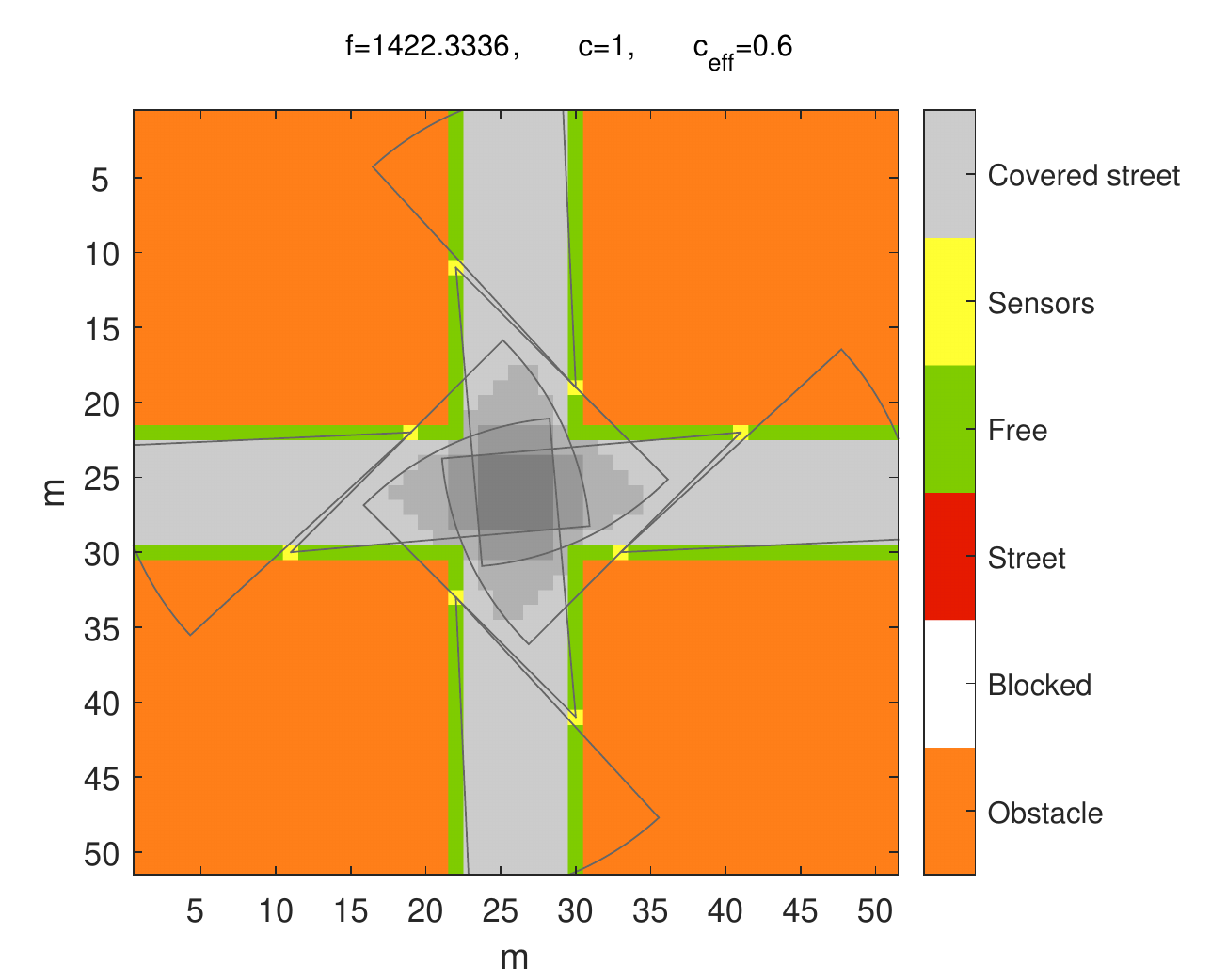}
\caption{Urban four-way intersection monitored by cameras ($r=20\m$, $\omega=40^{\circ}$), with a resolution of $\gs=1\m$. The central crossing zone of the intersection is assigned a double coverage priority. The road hosts two lanes.}
\label{fig:intersection}
\end{figure}

\begin{figure}[htb]
\centering
\includegraphics[width=.5\textwidth]{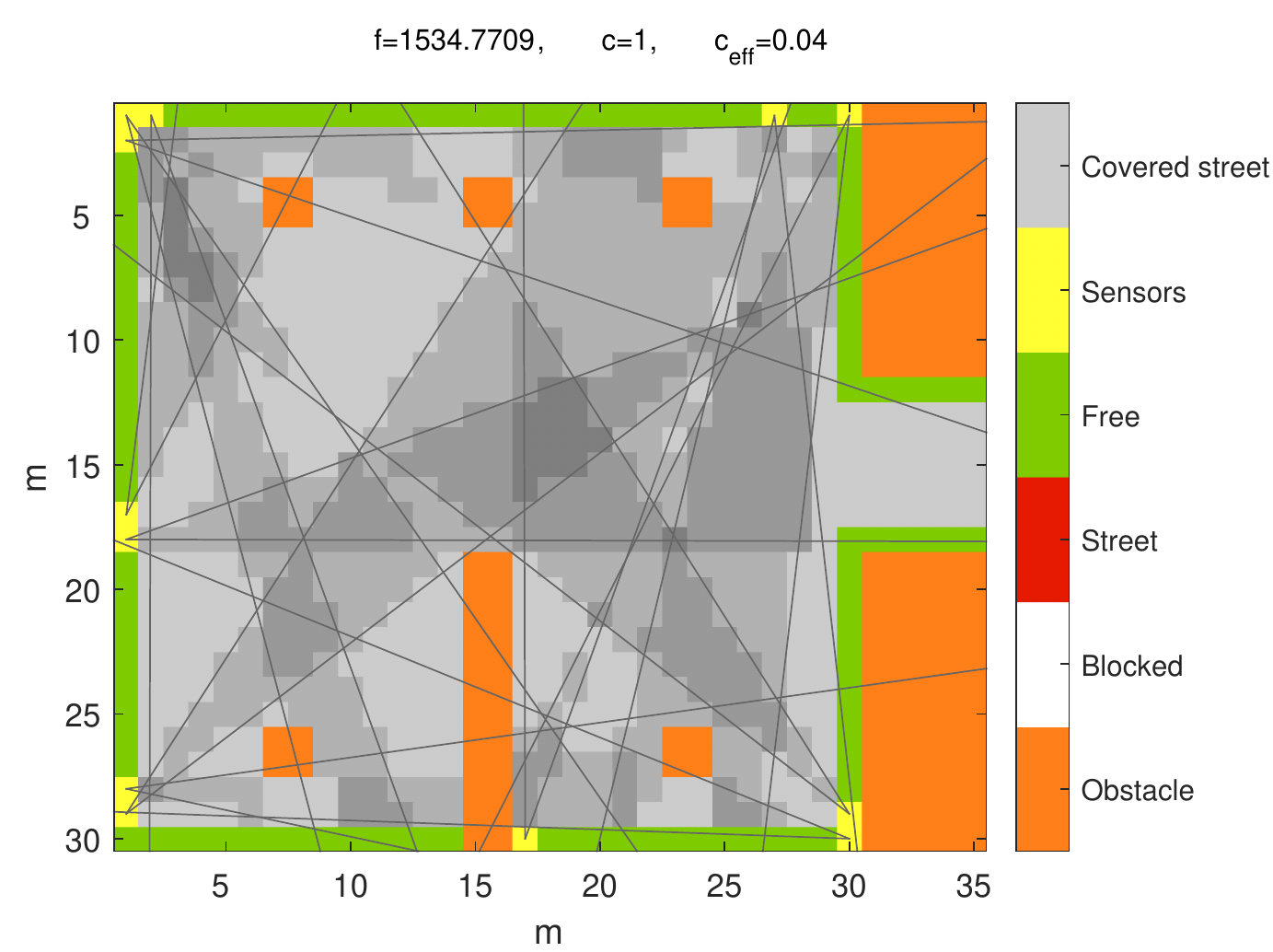}
\caption{Asymmetric parking space covered by high-precision radar sensors ($r=100\m$, $\omega=20^{\circ}$), grid length $\gs=1\m$. Pillars and walls block the line of sight.}
\label{fig:parking}
\end{figure}

\section{Results}
\label{sec:results}

\subsection{Parametrization and efficiency}
The various parameters controlling the genetic optimization procedure are interrelated in a complex way. Examining an intersection benchmark scenario, we find that this parametrization is well balanced by the choice given in Tab.~\ref{tab:GAparameter}. Since offspring generated by the sequential ranking described in Sec.~\ref{sec:crossover} is always more fit than any of the parents, the crossover rate is one.

\begin{table}[h]
\begin{center}
\begin{tabular}{ l r }
	\toprule
  Parameter & Assigned value  \\ 
	\midrule
	Population size $N$ & 150  \\ 
  Mutation rate $\pmut$ & 0.1  \\ 
  Crossover rate $\pcross$ & 1  \\ 
	Diversity $\pdiv$ & 0.3 \\ 
	\bottomrule
\end{tabular}
\end{center}
\caption{Parametrization of the genetic optimization algorithm, as determined for an intersection scenario of size $30\times30\m^2$, and a grid length of $\gs=1\m$.}
\label{tab:GAparameter}
\end{table}

To verify the efficiency of the method, we compare its outcome against the result of a simple greedy search approach. The latter places the sensors in a deterministic fashion, by picking subsequently the best possible option with respect to the gene ranking described in Sec.~\ref{sec:crossover}. Two selected scenarios similar to the ones in Sec.~\ref{sec:scenarios} are studied. For the parking garage setup, our algorithm achieves a $17\%$ higher efficiency $\ceff$ than the greedy search, at $\cov=1$. For a fully covered four-way intersection on the other hand, $\ceff$ increases drastically by $75\%$, mostly due to the additional symmetrization procedure. The computational complexity of the genetic optimization can be estimated to scale as $\mathcal{O}(N_{\text{gen, max}} \cdot N \cdot \Nroad \cdot N_{\text{sens, max}}^2)$, where $N_{\text{gen, max}}$ represents the maximum number of generations and $N_{\text{sens, max}}$ is the maximum number of sensors that are placed on the given map.

\subsection{Generic scenarios}
\label{sec:scenarios}

We illustrate our method for the following scenarios.

\subsubsection{Scenario 1 -- Multi-lane highway} 

Similar to the setup in \cite{Geissler2018}, the scenario of a four-lane motorway to be monitored by mid-range radar sensors is considered. 
In the case of right-hand side road traffic, the lanes to the right (the bottom in Fig.~\ref{fig:highway}) are typically more frequented, e.g. by slow and long vehicles, which leads to a higher probability of occlusions. We account for such traffic dynamics by modeling the rightmost lane as a highly opaque obstacle, and as a result, sensor positions on the opposite roadside are favored. 
Checking for translation-symmetric solutions, we find an optimal configuration in the form of staggered lineup (see Fig.~\ref{fig:highway}) with an inter-sensor distance of $46\m$.

\subsubsection{Scenario 2 -- Urban four-way intersection} 
We study the deployment of cameras with CCTV-typical sensing field characteristics \cite{Auckland2015} for an urban intersection, where building facades block the line of sight between adjacent legs of the intersection. The central crossing zone is assigned priority coverage.
Symmetrizing the outcome of the genetic optimization with respect to $C_4$ rotations significantly improves the fitness of the solution.
The most efficient configuration here forms a spiral-like pattern, with mutually interlocked sensor fields, see Fig.~\ref{fig:intersection}.

\subsubsection{Scenario 3 -- Parking garage} 

Third, we apply our optimization scheme to the scenario of an asymmetrically shaped parking garage containing various obstacles in the form of pillars and a separation wall. Monitoring in such a setup requires high accuracy and low sensitivity to illumination conditions, which is why we assume the deployment of high-precision radars with a narrow field of view \cite{Schneider2003} for high angular resolution.
The solution spends $12$ sensors to fully cover the garage, where favorable locations are for example found in the four corners (see Fig.~\ref{fig:parking}). Due the lack of symmetries and the long sensor range, the coverage efficiency is low.

\subsection{Scaling up}
\label{sec:scaling}
For increasing map sizes, the fast expansion of the search space challenges the optimization of infrastructure sensor placement, and the probability of finding efficient solutions reduces drastically.
To tackle this scaling issue, we fragment a large map into a set of much smaller, elementary units, which can be solved efficiently with the optimization method described above. Subsequently, the elements are reassembled to form the original environmental setup, where we develop a customized stitching algorithm to adjust the interfaces.

The algorithm consecutively joins pairs of road junctions and straight road segments, by checking the most favorable translational shift of a sensor ensemble populating a straight road segment, while maintaining the respective sensor inter-distances and angles. The degrees of freedom associated with mirror reflections of individual elements are taken into account by applying a mirror-specific greedy search to a set of randomly generated global configurations. 
Our method efficiently determines near-optimal global sensor placements even for large maps. 
In Fig.~\ref{fig:city}, we demonstrate the result of this procedure for an extended neighborhood in central Berlin, interfacing the solutions for four-way intersections and straight road segments that were derived in the above sections. The best configuration is picked here among a series of ten consecutive greedy searches. Our algorithm finds a solution with a high coverage efficiency $\ceff$, where all intersections have the same, clockwise chirality.

\section{Summary and conclusions}
\label{sec:summary}
We present a method to optimize sensor placement, even in complex environments, by combining the three steps of genetic evolution, a greedy local search, and symmetrization. This approach performs significantly better than a benchmark greedy search method for the studied scenarios, in particular the exploitation of underlying map symmetries improves the fitness of a network configuration.
Dynamic obstacles on the road are considered by either assigning priorities or defining transparent sensing barriers. This feature is key for realistic scenarios, as it can be used for example to mimic expected road traffic patterns. 
Eventually, we demonstrate an approach to efficiently optimize sensor placement on large maps, by interfacing sensor structures that are found to be optimal on a small scale. Different sensor types can be studied here as long as all sensors are homogeneous.

\section*{Acknowledgment}
We thank Alexander Unnervik and Michael Paulitsch for fruitful discussions. This work was funded by the German Federal Ministry of Transport, Building and Urban Development (BMVI) within the project KoRA9 (grant No. 16AVF1032A).

\bibliographystyle{IEEEtran}
\bibliography{SensorBib}

\begin{thebibliography}{10}
\providecommand{\url}[1]{#1}
\csname url@samestyle\endcsname
\providecommand{\newblock}{\relax}
\providecommand{\bibinfo}[2]{#2}
\providecommand{\BIBentrySTDinterwordspacing}{\spaceskip=0pt\relax}
\providecommand{\BIBentryALTinterwordstretchfactor}{4}
\providecommand{\BIBentryALTinterwordspacing}{\spaceskip=\fontdimen2\font plus
\BIBentryALTinterwordstretchfactor\fontdimen3\font minus
  \fontdimen4\font\relax}
\providecommand{\BIBforeignlanguage}[2]{{%
\expandafter\ifx\csname l@#1\endcsname\relax
\typeout{** WARNING: IEEEtran.bst: No hyphenation pattern has been}%
\typeout{** loaded for the language `#1'. Using the pattern for}%
\typeout{** the default language instead.}%
\else
\language=\csname l@#1\endcsname
\fi
#2}}
\providecommand{\BIBdecl}{\relax}
\BIBdecl

\bibitem{Fuerstenberg2005}
K.~Fuerstenberg, ``{Advanced Intersection Safety - The EC project INTERSAFE
  (IV'06)},'' \emph{IV'06 IEEE Intelligent Vehicles symposium, Tokio, Japan},
  pp. 89--93, 2005.

\bibitem{Leader2004}
S.~Leader, ``{Telecommunications Handbook for Transportation Professionals -
  The Basics of Telecommunications},'' Federal Highway Administration, Tech.
  Rep., 2004.

\bibitem{Fayta2015}
\BIBentryALTinterwordspacing
M.~Fayta, ``{Connected Vehicles and Mcity},'' 2015. [Online]. Available:
  \url{http://itswisconsin.org/wp-uploads/2017/07/2015-Forum-Fayta.pdf}
\BIBentrySTDinterwordspacing

\bibitem{Quek2016}
\BIBentryALTinterwordspacing
A.~Quek, ``{Singapore Autonomous Vehicle Initiative ( SAVI )},'' 2016.
  [Online]. Available:
  \url{https://www.itu.int/en/ITU-T/extcoop/cits/Documents/Workshop-201707-Singapore/010
  - Alan-Quek-Singapore Autonomous Vehicle Initiative (SAVI).pdf}
\BIBentrySTDinterwordspacing

\bibitem{Audi2016}
\BIBentryALTinterwordspacing
Audi, ``{Audi launches first Vehicle-to-Infrastructure (V2I) technology in the
  U.S. starting in Las Vegas},'' 2016. [Online]. Available: \url{Audi launches
  first Vehicle-to-Infrastructure (V2I) technology in the U.S. starting in Las
  Vegas}
\BIBentrySTDinterwordspacing

\bibitem{Safespot2006}
\BIBentryALTinterwordspacing
``{Safespot - Cooperative vehicles and road infrastructure for road safety}.''
  [Online]. Available: \url{http://www.safespot-eu.org/deliverables.html}
\BIBentrySTDinterwordspacing

\bibitem{BMVI2019}
\BIBentryALTinterwordspacing
BMVI, ``{Kooperative Radarsensoren f{\"{u}}r das digitale Testfeld A9 -
  KoRA9},'' 2019. [Online]. Available:
  \url{https://www.bmvi.de/SharedDocs/DE/Artikel/DG/AVF-projekte/KoRA9.html}
\BIBentrySTDinterwordspacing

\bibitem{GM2019}
{Google Maps}, ``{Berlin Gendarmenmarkt}.''

\bibitem{Zhu2012}
C.~Zhu, C.~Zheng, L.~Shu, and G.~Han, ``{A survey on coverage and connectivity
  issues in wireless sensor networks},'' \emph{Journal of Network and Computer
  Applications}, vol.~35, no.~2, pp. 619--632, 2012.

\bibitem{AmacGuvensan2011}
M.~{Amac Guvensan} and A.~{Gokhan Yavuz}, ``{On coverage issues in directional
  sensor networks: A survey},'' \emph{Ad Hoc Networks}, vol.~9, no.~7, pp.
  1238--1255, 2011.

\bibitem{VanDenHengel2009}
A.~{Van Den Hengel}, R.~Hill, B.~Ward, A.~Cichowski, H.~Detmold, C.~Madden,
  A.~Dick, and J.~Bastian, ``{Automatic camera placement for large scale
  surveillance networks},'' \emph{2009 Workshop on Applications of Computer
  Vision, WACV 2009}, pp. 1--6, 2009.

\bibitem{Indu2009}
S.~Indu, A.~Bhattacharyya, N.~R. Mittal, and S.~Chaudhury, ``{Optimal sensor
  placement for surveillance of large spaces},'' \emph{2009 3rd ACM/IEEE
  International Conference on Distributed Smart Cameras, ICDSC 2009}, pp. 1--8,
  2009.

\bibitem{Yoon2013}
{Yourim Yoon} and {Yong-Hyuk Kim}, ``{An Efficient Genetic Algorithm for
  Maximum Coverage Deployment in Wireless Sensor Networks},'' \emph{IEEE
  Transactions on Cybernetics}, vol.~43, no.~5, pp. 1473--1483, 2013.

\bibitem{Zhang2016}
G.~Zhang, S.~You, J.~Ren, D.~Li, and L.~Wang, ``{Local coverage optimization
  strategy based on voronoi for directional sensor networks},'' \emph{Sensors
  (Switzerland)}, vol.~16, no.~12, pp. 1--15, 2016.

\bibitem{Cheng2010}
X.~Cheng, P.~Liu, Z.~Chen, H.~Wu, and X.~Fan, ``{The Optimal Sensing Coverage
  for Road Surveillance.}'' \emph{Engineering}, vol.~2, no.~4, pp. 318--327,
  2010.

\bibitem{Field2006}
R.~V. Field and M.~Grigoriu, ``{Optimal design of sensor networks for vehicle
  detection, classification, and monitoring},'' \emph{Probabilistic Engineering
  Mechanics}, vol.~21, no.~4, pp. 305--316, 2006.

\bibitem{Tao2006}
D.~Tao, H.~Ma, and L.~Liu, ``{Coverage-enhancing algorithm for directional
  sensor networks},'' \emph{Lecture Notes in Computer Science (including
  subseries Lecture Notes in Artificial Intelligence and Lecture Notes in
  Bioinformatics)}, vol. 4325 LNCS, pp. 256--267, 2006.

\bibitem{Cheng2007}
W.~Cheng, S.~Li, X.~Liao, S.~Changxiang, and H.~Chen, ``{Maximal coverage
  scheduling in randomly deployed directional sensor networks},''
  \emph{Proceedings of the International Conference on Parallel Processing
  Workshops}, 2007.

\bibitem{Zhao2009}
J.~Zhao and J.~C. Zeng, ``{An electrostatic field-based coverage-enhancing
  algorithm for wireless multimedia sensor networks},'' \emph{Proceedings - 5th
  International Conference on Wireless Communications, Networking and Mobile
  Computing, WiCOM 2009}, 2009.

\bibitem{Holland75}
J.~H. Holland, \emph{{Adaptation in natural and artificial systems: an
  introductory analysis}}.\hskip 1em plus 0.5em minus 0.4em\relax Ann Arbor,
  Mich.: Univ. of Michigan Press, 1975.

\bibitem{Ferrari2006}
S.~Ferrari, ``{Track Coverage in Sensor Networks},'' \emph{Proceedings of the
  2006 American Control Conference}, 2006.

\bibitem{Srinivas94}
M.~Srinivas and L.~M. Patnaik, ``{Genetic Algorithms : A Survey},'' \emph{IEEE
  Computer}, vol.~27, no.~6, pp. 17--26, 1994.

\bibitem{Deb2001}
K.~Deb, \emph{{Multi-Objective Optimization Using Evolutionary
  Algorithms}}.\hskip 1em plus 0.5em minus 0.4em\relax John Wiley {\&} Sons,
  Ltd, 2001, vol.~16.

\bibitem{Geissler2018}
F.~Geissler, S.~Kohnert, and R.~Stolle, ``{Designing a Roadside Sensor
  Infrastructure to Support Automated Driving},'' in \emph{2018 IEEE
  International Conference on Intelligent Transportation Systems (ITSC)}, 2018.

\bibitem{Auckland2015}
\BIBentryALTinterwordspacing
``{CCTV camera data},'' 2015. [Online]. Available:
  \url{http://www.cctv-auckland.com/cctv-camera-lens-distance-angles-and-coverages.html}
\BIBentrySTDinterwordspacing

\bibitem{Schneider2003}
R.~Schneider and J.~Wenger, ``{High resolution radar for automobile
  applications},'' \emph{Advances in Radio Science}, vol.~1, pp. 105--111,
  2003.

\end{thebibliography}

\end{document}